\documentclass[prl,twocolumn,floatfix,superscript address]{revtex4}

\usepackage{graphicx,epic,eepic,epsfig,amsmath,latexsym,amssymb,verbatim,subfigure,color}
\usepackage{verbatim,bbm}

\usepackage{theorem}

\newtheorem{definition}{Definition}

\newtheorem{lemma}[definition]{Lemma}

\newtheorem{theorem}[definition]{Theorem}

\def\amplitude{bit }
\def\squareforqed{\hbox{\rlap{$\sqcap$}$\sqcup$}}
\def\qed{\ifmmode\squareforqed\else{\unskip\nobreak\hfil
\penalty50\hskip1em\null\nobreak\hfil\squareforqed
\parfillskip=0pt\finalhyphendemerits=0\endgraf}\fi}
\def\endenv{\ifmmode\;\else{\unskip\nobreak\hfil
\penalty50\hskip1em\null\nobreak\hfil\;
\parfillskip=0pt\finalhyphendemerits=0\endgraf}\fi}
\newenvironment{proof}{\noindent \textbf{{Proof~} }}{\qed}

\mathchardef\ordinarycolon\mathcode`\:
\mathcode`\:=\string"8000
\def\vcentcolon{\mathrel{\mathop\ordinarycolon}}
\begingroup \catcode`\:=\active
  \lowercase{\endgroup
  \let :\vcentcolon
  }

\newcommand{\nc}{\newcommand}
\nc{\rnc}{\renewcommand}
\nc{\beq}{\begin{equation}}
\nc{\eeq}{{\end{equation}}}
\nc{\beqa}{\begin{eqnarray}}
\nc{\eeqa}{\end{eqnarray}}
\nc{\lbar}[1]{\overline{#1}}
\nc{\bra}[1]{\langle#1|}
\nc{\ket}[1]{|#1\rangle}
\nc{\ketbra}[2]{|#1\rangle\!\langle#2|}
\nc{\braket}[2]{\langle#1|#2\rangle}
\nc{\proj}[1]{| #1\rangle\!\langle #1 |}
\nc{\avg}[1]{\langle#1\rangle}
\rnc{\max}{\operatorname{max}}
\nc{\Rank}{\operatorname{Rank}}
\nc{\smfrac}[2]{\mbox{$\frac{#1}{#2}$}}
\nc{\Tr}{\operatorname{Tr}}
\nc{\id}{\operatorname{id}}
\nc{\1}{\openone}
\nc{\ox}{\otimes}
\nc{\dg}{\dagger}
\nc{\dn}{\downarrow}
\nc{\cA}{{\cal A}}
\nc{\cB}{{\cal B}}
\nc{\cC}{{\cal C}}
\nc{\cD}{{\cal D}}
\nc{\cE}{{\cal E}}
\nc{\cF}{{\cal F}}
\nc{\cG}{{\cal G}}
\nc{\cH}{{\cal H}}
\nc{\cI}{{\cal I}}
\nc{\cJ}{{\cal J}}
\nc{\cK}{{\cal K}}
\nc{\cL}{{\cal L}}
\nc{\cM}{{\cal M}}
\nc{\cN}{{\cal N}}
\nc{\cO}{{\cal O}}
\nc{\cP}{{\cal P}}
\nc{\cR}{{\cal R}}
\nc{\cS}{{\cal S}}
\nc{\cT}{{\cal T}}
\nc{\cX}{{\cal X}}
\nc{\cZ}{{\cal Z}}
\nc{\supp}{{\operatorname{supp}}}
\nc{\var}{\operatorname{var}}
\nc{\rar}{\rightarrow}
\nc{\lrar}{\longrightarrow}
\nc{\polylog}{\operatorname{polylog}}

\def\ph{\varphi}

\nc{\RR}{{{\mathbb R}}}
\nc{\CC}{{{\mathbb C}}}
\nc{\FF}{{{\mathbb F}}}
\nc{\NN}{{{\mathbb N}}}
\nc{\ZZ}{{{\mathbb Z}}}
\nc{\PP}{{{\mathbb P}}}
\nc{\QQ}{{{\mathbb Q}}}
\nc{\UU}{{{\mathbb U}}}
\nc{\EE}{{{\mathbb E}}}

\nc{\be}{\begin{equation}}
\nc{\ee}{{\end{equation}}}
\nc{\bea}{\begin{eqnarray}}
\nc{\eea}{\end{eqnarray}}
\nc{\<}{\langle}
\rnc{\>}{\rangle}
\nc{\Hom}[2]{\mbox{Hom}(\CC^{#1},\CC^{#2})}
\nc{\rU}{\mbox{U}}

\begin{document}
\author{Graeme Smith}
\affiliation{Institute for Quantum Information, Caltech 107--81,
    Pasadena, CA 91125, USA}
\affiliation{IBM T.J. Watson Research Center, Yorktown Heights, NY 10598, smolin@watson.ibm.com}
\author{Joseph M.~Renes}
\affiliation{Institut f\"ur Angewandte Physik, Technische Universit\"at Darmstadt, 64289 Darmstadt, Germany}
\author{John A.~Smolin}
\affiliation{IBM T.J. Watson Research Center, Yorktown Heights, NY 10598, smolin@watson.ibm.com}

\title{Structured codes improve the Bennett-Brassard-84 quantum key rate}

\begin{abstract}
A central goal in information theory and cryptography is finding
simple characterizations of optimal communication rates subject to
various restrictions and security requirements.  Ideally, the optimal
key rate for a quantum key distribution (QKD) protocol would be given by
{\em single-letter formula} involving a simple optimization over a single
use of an effective channel.  We explore the possibility of
such a formula for one of the simplest and most widely
used QKD protocols---Bennett-Brassard-84
(BB84) with one way classical post-processing.  
We show that a conjectured single-letter key-rate formula is false,
uncovering a deep ignorance about asymptotically good private
codes and  pointing towards unfortunate complications in the theory of QKD.  
These complications are
not without benefit---with added complexity comes better key rates than
previously thought possible.  We improve the threshold for secure key
generation from a bit error rate of 0.124 to 0.129.
\end{abstract}

\date{July 4, 2006}
\maketitle

Quantum key distribution (QKD) allows two parties using public
channels to remotely establish a secret key whose security is not
predicated on the difficulty of some computational task.  Rather, the
security of the key generated by a QKD protocol depends only on
fundamental laws of physics.  As a result there has been an enormous
amount of work on practical and theoretical aspects of QKD, and a
corresponding rapid progress in both \cite{GisinReview02}.

The first QKD protocol was proposed by Bennett and Brassard in 1984~\cite{BB84}, and 
like all QKD schemes, it is based on the tradeoff between information gain 
and disturbance in quantum mechanics.  To establish a bit of raw key, 
the sender (Alice) encodes a random bit into one of two conjugate bases ($X$ or $Z$),
chosen at random, and transmits it to a receiver (Bob).  Bob measures in either
the $X$ or $Z$ basis, also chosen at random.  After generating a large number of 
bits(say, $2n$), Alice and Bob can sift out the bits for which they both chose the same 
basis by public discussion, leaving roughly $n$ bits.

Alice then randomly permutes her remaining bits and announces the
permutation to Bob, after which they perform parameter estimation by
comparing a small fraction of their bits to find the error rate
of the sifted key.  If the fraction $p$ of bits on which they
disagree is sufficiently small, they proceed with information
reconciliation and privacy amplification to finally arrive at a secret
key. The essence of the protocol is that if an eavesdropper Eve, who
is assumed to have control of the quantum channel, examines the
signals in order to determine the key, she will necessarily cause some
disturbance which manifests itself as errors in the sifted key. Thus
$p$ also characterizes how much Eve could have learned about the key.
 
An important property of any QKD protocol is the amount of noise that
can be tolerated without compromising the privacy of the resulting
key, the amount of noise at which the protocol aborts. The
entanglement-based security proof of Shor and
Preskill~\cite{ShorPreskill00} showed that BB84 can be used to
generate private key for detected bit error rates as high as $p\approx
0.11$, basically by showing there exist Calderbank-Shor-Steane
(CSS)~\cite{Steane96,CalderShor96} codes correcting noise up to this
level.  Remarkably, it was recently found~\cite{KGR05,RGK05} that this
can be improved to $p\approx 0.124$ if Alice adds independent noise to
her sifted key before performing the distillation steps, which has
been conjectured to be optimal among all one-way key distillation protocols \cite{RGK05}.  
The key rates of
\cite{KGR05} come from evaluating a {\em single-letter} key rate for
an effective state found by Devetak and Winter in \cite{D03}, and indeed the 
$0.124$ threshold of \cite{KGR05,RGK05} is the optimal threshold for this 
single-letter formula \footnote{The optimality of the $0.124$ threshold for the single-letter
Devetak-Winter formula has not been proven in the literature, but is easily 
verified by numerical optimization.}.  
If these rates
were optimal among {\em all} protocols, it would indicate a 
single-letter formula for one-way QKD key-rates, providing a dramatic simplification in the
theory of quantum key distribution protocols.

We will show that $p\approx 0.124$
is {\em not} optimal, and the threshold is at least $p\approx
0.129$. We increase the threshold by finding improved error correcting
codes for the information reconciliation phase. The technique is
analogous to those of \cite{SS96,DSS98,SmithSmo0506}, which use
degenerate CSS codes to achieve higher quantum capacities than are
achievable by the single-letter formula for quantum capacity arising
from random stabilizer codes.  Though the true maximization needed for
the multi-letter capacity formula in \cite{D03} remains out of reach, we are able to
evaluate rates for particular multi-letter inputs which achieve
higher key rates than the single-letter maximum.  While this is suggestive, 
we emphasize that our results to not necessarily rule out a single-letter formula 
for the one-way key-rate.  We have shown that the single-letter Devetak-Winter 
formula does not give the one-way distillable key, but this does not preclude the existence of
some other single-letter optimization problem that gives the optimal key rate.  

Taken together, our information reconciliation and privacy
amplification steps can be described by a highly degenerate CSS code.
A quantum code is called degenerate if its syndrome does not uniquely
identify the errors which it corrects.  This is a uniquely quantum
effect---there is no such thing as a degenerate classical code--and
all such codes involve entanglement.  It appears remarkable then that
degeneracy should help in the classical processing task of key
distillation.  Moreover, Alice and Bob need not perform any
multi-particle quantum operations even in our improved protocol.
The resolution is that Eve's best attacks involve entanglement, 
and degeneracy will make this work against her.

Degenerate codes have been used for QKD before;
specifically, to improve the threshold of the six-state protocol
from $0.126$ to $0.127$ \cite{Lo01}.  However, this protocol did
not involve noisy processing, and in fact a better threshold was
obtained for the six-state protocol by~\cite{KGR05,RGK05}.  Our result
combines degenerate codes with noisy processing, leading to
an advantage over either one alone.

{\em Analytic key rate expression---}To determine the secret key
rate of the modified protocol, we follow~\cite{KGR05,RGK05, RenThesis}. 
First, the prepare \&
measure protocol can be converted to an equivalent scheme in which
Alice prepares the maximally-entangled state $\ket{\Phi^+}^{\otimes
mn}_{AB}$ and sends half to Bob. Each party then randomly and
independently measures either $X$ or $Z$ on each signal, saving the
outcomes for use in parameter estimation and key generation. They
discard the outcomes where their basis choice did not agree,
and denoting the remaining outcomes $K_A$ and $K_B$ it
follows from Corollary 6.5.2 of \cite{RenThesis} that for any
$m$-bit processing step $K_A^m\rightarrow U$ and $U\rightarrow V$ it
is possible to use standard ({\em i.e.}, unstructured, random) error
correction and privacy amplification to distill secret key at rate
\begin{equation}
\label{Eq:RenRateMany}
r=\frac{1}{m}\inf_{\sigma_{A B}\in \Gamma_{p}} \big[S(U|VE^m) - S(U|VK_B^m)\big],
\end{equation}
evaluated on the state generated by performing the processing on
$\sigma_{AB}^{\ox m}$, and where $\Gamma_{p}$ is the set of single
pair Bell-diagonal states $\sigma_{AB}$ passing the parameter
estimation phase of the protocol and $E^m$ is the purification of
$\sigma_{AB}^{\ox m}$, which we must assume belongs to Eve.  $S(\rho)
= -\Tr \rho \log \rho$ is the von Neumann entropy.  This expression is
similar to what was found in~\cite{KGR05, RGK05}, with the additional
feature that it includes blockwise processing.  Since the $X$ and $Z$
bases are randomly used to create the sifted key, the error estimation
provides an estimate of the bit- and phase-flip noise rates, so that
the allowable $\sigma_{AB}$ are of the form
$\sigma_{AB}=(1+t-2p)\ketbra{\Phi^+}{\Phi^+}+(p-t)(\ketbra{\Phi^-}{\Phi^-}+\ketbra{\Psi^+}{\Psi^+})+t\ketbra{\Psi^-}{\Psi^-}$
for $t\in[0,p]$.

Below, we choose a particular $K_A^m\rightarrow U\rightarrow V$ for which Eq.~(\ref{Eq:RenRateMany}) 
outperforms all previously known protocols for large $p$.
The measurements leading to $K_A$ and $K_B$ will be the same as for the usual BB84 protocol, 
with the processing step chosen as follows.  For each 
$m$ bit block of $K_A$, $(x_1,x_2,\dots,x_m)$, Alice independently flips each bit 
with probability $q$, resulting in $\tilde{\mathbf x} = (\tilde{x}_1,\dots,\tilde{x}_m)$.  
She then computes $U = (\tilde{x}_1,\tilde{x}_1\oplus \tilde{x}_2, \dots,\tilde{x}_1\oplus \tilde{x}_m)$ and sends
$V = (\tilde{x}_1\oplus \tilde{x}_2, \dots,\tilde{x}_1\oplus \tilde{x}_m)$ to Bob, 
after which they do error correction and privacy amplification
as usual.  The key rate they achieve is given by the following theorem.

\begin{theorem} \label{Thm:KeyRate}
The key rate achieved using the processing $\mathbf{x}\rightarrow U\rightarrow V$   with  
$U =(\tilde{x}_1,\tilde{x}_1\oplus \tilde{x}_2, \dots,\tilde{x}_1\oplus \tilde{x}_m)$,
$V = (\tilde{x}_1\oplus \tilde{x}_2, \dots,\tilde{x}_1\oplus \tilde{x}_m)$, where $\tilde{\mathbf x} = {\mathbf x}\oplus {\mathbf f}$ and $\mathbf f$ is a
string of independent 0-1 random variables, each with probability $q$ of being $1$, is given by
\begin{eqnarray}\label{Eq:OverallKeyRate}
r&=&\frac{1}{m}\Bigg(1 - \sum_{\mathbf s}P^{\tilde{p}}_{m}(\mathbf s)H\left(P^{\tilde{p}}_{m}(u|\mathbf s)\right) +
m S(\rho_{p,q}) \nonumber\\
&&\phantom{\frac{1}{m}\Bigg(1}- S\left(\frac{1}{2}\rho_{p,q}^{\ox m}+\frac{1}{2}Z^{\ox m} \rho_{p,q}^{\ox m}Z^{\ox m}\right) \Bigg).
\end{eqnarray}
Here $\rho_{p,q} = (1-q)\proj{\ph_{+}} +q\proj{\ph_{-}} $ with 
$\ket{\ph_{\pm}} = \sqrt{1-p}\ket{0}\pm\sqrt{p}\ket{1}$, $\tilde{p} = p(1-q)+q(1-p)$, while
$P^{\tilde{p}}_{m}(u,\mathbf s)$ is defined in Lemma \ref{Lem:RepRates}.  
The entropy $H$ of a classical probability distribution $P$ is given by
$H(P) = -\sum_l P_l \log P_l$.
\end{theorem}     

We proceed by noting that in the entanglement picture, our processing step is equivalent to Alice first adding independent
\amplitude errors to her halves of the noisy EPR pairs, measuring the stabilizers of an $m$ qubit repetition code, and then sending her
syndrome outcomes to Bob.  We apply the following lemma, which follows from \cite{SmithSmo0506}.

\begin{lemma}\label{Lem:RepRates}
The $m$ qubit repetition code with stabilizers $Z_1Z_2,\dots,Z_1Z_m$ maps the error $X^{\mathbf u}Z^{\mathbf v}$ to 
the logical error $X^{u_1}Z^{\oplus_{l=1}^m v_l}$ and syndrome ${\mathbf s} = (u_1\oplus u_2,\dots u_1\oplus u_m)$.  When used
to correct independent \amplitude errors of probability $p$, the probability
of a logical \amplitude error $u$ and syndrome $\mathbf s$ is given by 
\begin{equation}\label{Eq:Lemma2}
P^{p}_{m}(u,\mathbf s) = \left(p^{m-s}(1-p)^s\right)^u\left(p^{s}(1-p)^{m-s}\right)^{1-u},
\end{equation}
for $s = |\mathbf s|$.
\end{lemma}

\begin{proof}{\bf of Theorem \ref{Thm:KeyRate}}
To evaluate Eq.~(\ref{Eq:RenRateMany}), first let
\begin{equation}
\sigma^{\ox m}_{AB} = \sum_{\mathbf u,\mathbf v}p_{\mathbf u \mathbf v} X^{\mathbf u}_B Z^{\mathbf v}_B
\big[\proj{\Phi^+}\big]_{AB}^{\ox m}Z^{\mathbf v}_B X^{\mathbf u}_B,
\end{equation}
with $p_{\mathbf u\mathbf v}$ such that 
$p_{\mathbf u}=\sum_{\mathbf v}p_{\mathbf u \mathbf v}=p^{|{\mathbf u}|}(1-p)^{m-|{\mathbf u}|}$, 
for measured bit error rate $p$, and similarly for $p_{\mathbf v}$.

Alice adds independent noise at error rate $q$ to the $A$ register, so the state of the Alice-Bob-Eve system can be described as
\begin{equation}
\sum_{\mathbf{u, v, f}} \sqrt{p_{\mathbf u \mathbf v}q_{\mathbf f}} \, 
\ket{\mathbf f}_{A^\prime} X^{\mathbf u}_B Z^{\mathbf v}_B X^{\mathbf f}_B 
\ket{\Phi^+}_{AB}^{\ox m}\ket{\mathbf u}_{E_1}\ket{\mathbf v}_{E_2},
\end{equation}
where we have used the fact that $X_A\ox I \ket{\Phi^+}_{AB} = I \ox X_B \ket{\Phi^+}_{AB}$.
Note that Eve's system is determined by the fact that in the worst case she holds the purification of the state after it
emerges from the channel. However, she does not hold the purification of the noise Alice adds.

Alice and Bob then measure the stabilizers of the $m$-qubit repetition code 
($Z_1 Z_2,\dots Z_1 Z_m$) and Alice sends her outcomes to Bob.  
This is equivalent to having Bob defers his measurement until he receives Alice's message and then 
coherently correcting his key bit, which we will consider here.
Renaming Bob's $m-1$ syndrome qubits system $B'$, 
the state they'll share in this case is
\begin{eqnarray}\label{Eq:StartState}
&\sum_{\mathbf{u, v, f}}&\sqrt{p_{\mathbf u\mathbf v}q_{\mathbf f}} \, 
\ket{\mathbf f}_{A^\prime} X^{u_1\oplus f_1}_B Z^{\oplus_{l=1}^m v_l}_B
\ket{\Phi^{+}}_{AB}\\
&&\otimes \ket{\mathbf{s}_{\mathbf{u},\mathbf{f}}}_{B^\prime} 
\ket{\mathbf u}_{E_1}Z^{\mathbf f}_{E_2}\ket{\mathbf v}_{E_2},\nonumber
\end{eqnarray}
where $\mathbf{s}_{\mathbf{u},\mathbf{f}}$ is an $(m-1)$-bit string labeling the basis states of $B^\prime$ 
whose $j$th bit is $(\mathbf{s}_{\mathbf{u},\mathbf{f}})_j=u_1\oplus u_{j+1}\oplus f_1\oplus f_{j+1}$.
Note that the $Z^{\mathbf f}$ acting on Eve's second system comes from the 
commutation of $Z^{\mathbf v}_B$ and $X^{\mathbf f}_B$.

Getting rid of the $A^\prime$ system (but keeping it from Eve), we now
let Alice and Bob measure systems $A$ and $BB^\prime$ in the
computational basis, respectively.  According to
Eq.~(\ref{Eq:RenRateMany}), the difference of conditional entropies
for the resulting state will give us the key rate.  This will be
simpler to analyze by first rewriting the lower bound as
\begin{equation}\label{Eq:RenRateRep}
r \geq \frac{1}{m}\inf_{\sigma_{AB}\in \Gamma_{p}} I(A;BB^\prime)-I(A;E).
\end{equation}
$I(A;BB^\prime)$ is the mutual information ($I(X;Y)=S(X)+S(Y)-S(XY)$) 
of 
$\rho_{ABB'}=\frac{1}{2}\sum_{x=0}^1\proj{x}_A \ox \rho_{B^\prime B}^{x}$,
where
\begin{eqnarray}
\rho_{B^\prime B}^{x} &=& 
\sum_{\mathbf f}\sum_{\mathbf u}q_{\mathbf f}p_{\mathbf u}\,\proj{x{+}f_1{+}u_1}_B\ox 
\proj{\mathbf{s}_{\mathbf{u},\mathbf{f}}}\nonumber\\
& = & \sum_{\mathbf s}P^{\tilde{p}}_{m}(\mathbf s)\! \sum_{u=0}^1 P^{\tilde{p}}_{m}(u|\mathbf s)\proj{x{+}u}_B\ox \proj{\mathbf s}_{B^\prime},\nonumber
\end{eqnarray}
and the $P^{\tilde{p}}_{m}(u,\mathbf s)$ are given by Lemma
\ref{Lem:RepRates}.  Thus, the mutual information,
$I(A;BB^\prime)$, is exactly $1 - \sum_{\mathbf
s}P^{\tilde{p}}_{m}(\mathbf s)H(P^{\tilde{p}}_{m}(u|\mathbf s))$.
Notice that this term only depends on $p_{\mathbf u}$, which is
determined by the parameter estimation phase, so it will be the same
for all ${\sigma_{AB}\in \Gamma_{p}}$.

Turning to the second term in Eq.~(\ref{Eq:RenRateRep}), we want to find the mutual information of the Alice-Eve system,
$
\rho_{AE_1E_2}  =  \frac{1}{2}\sum_{x=0}^1\proj{x}_A \ox \rho_{E_1E_2}^{x},
$
where
\begin{widetext}
\begin{eqnarray}
\rho_{E_1E_2}^{x} & = & \left(Z_{E_2}^{\ox m}\right)^x\left(\sum_{\mathbf u ,\mathbf v_1,\mathbf v_2,\mathbf f} 
q_{\mathbf f}\sqrt{p_{\mathbf{u}|\mathbf{v}_1}p_{\mathbf{u}|\mathbf{v}_2}}\,\proj{\mathbf u}_{E_1}
\ox \sqrt{p_{\mathbf{v}_1}p_{\mathbf{v}_2}}\,Z^{\mathbf f}\ketbra{\mathbf v_1}{\mathbf v_2}_{E_2}Z^{\mathbf f}  \right) \left(Z_{E_2}^{\ox m}\right)^x.
\end{eqnarray}
\end{widetext}
Note that the $(Z_{E_2}^{\otimes m})^x$ comes from the action of $Z^{\oplus_{l=1}^m v_l}$ on $B$.
When \amplitude and phase errors are independent, this expression can be further simplified. Defining 
$\mu=\sum_\mathbf{u} p_{\mathbf{u}}\proj{\mathbf{u}}$ and
$\rho_{p,q} = (1{-}q)\proj{\ph_{+}}{+}q\proj{\ph_{-}} $ with $\ket{\ph_{\pm}} = \sqrt{1{-}p}\ket{0}{\pm}\sqrt{p}\ket{1}$, we can write
\begin{equation}
\rho_{E_1,E_2}^x=\mu_{E_1}\otimes \left(Z_{E_2}^{\ox m}\right)^x
\left[\rho_{p,q}^{\ox m}\right]_{E_2} \left(Z_{E_2}^{\ox m}\right)^x.
\end{equation}
Actually, we have to maximize $I(A;E_1 E_2)$ over all 
$p_{\mathbf u \mathbf v}$ corresponding to states in ${\sigma_{AB}\in \Gamma_{p}}$, but 
the largest value is attained for independent phase 
and \amplitude errors.  This means that Eve's optimal attack on the protocol will be to choose $\sigma_{AB}\in\Gamma_p$ with $t=p^2$. 
In particular, if Eve starts with the independent 
$\mathbf u, \mathbf v$ state, by tracing out the $E_1$ system and using the isometry
\begin{equation}
U = \sum_{\mathbf v, \mathbf u}\sqrt{p_{\mathbf u|\mathbf v}}\ket{\mathbf u}_{E_3}\ket{\mathbf v}_{E_2}\bra{\mathbf v}_{E_2},
\end{equation}
then completely dephasing $E_3$, she can construct a 
$\rho_{AE_2E_3}$ with the same mutual information as if the errors were distributed 
according to $p_{\mathbf u|\mathbf v}p_{\mathbf v}$.  Since mutual information 
cannot be increased by local operations, the independent noise state must have 
the largest value.  Moreover, as the $E_1$ system is uncorrelated with $AE_2$, 
$I(A;E)$ can be easily computed, yielding
\begin{equation}
\nonumber
I(A;E) = S\left(\frac{1}{2}\rho_{p,q}^{\ox m}+\frac{1}{2}Z^{\ox m} \rho_{p,q}^{\ox m}Z^{\ox m}\right) - m S(\rho_{p,q}).
\end{equation}
Taking the difference between $I(A;BB^\prime)$ and $I(A;E)$,
keeping in mind we must send $m$ qubits for each $m$-block, leads to the overall key rate of Eq.~(\ref{Eq:OverallKeyRate}).
\end{proof}

{\em Numerical key rates---}We now
evaluate Eq.~(\ref{Eq:OverallKeyRate}) for particular
$p$, $q$, and $m$.  $S(\rho_{p,q})$ is
easily calculated and the second term can be evaluated efficiently
via Eq.~(\ref{Eq:Lemma2}).  The most difficult term is
$S\left(\frac{1}{2}\rho_{p,q}^{\ox m}{+}\frac{1}{2}Z^{\ox m}
\rho_{p,q}^{\ox m}Z^{\ox m}\right) $, but it can be handled as
follows. Due to the permutation-invariance of the state
$\rho_{p,q}^{\ox m}$, it is compactly expressed as a direct sum over
the $SU(2)$ irreducible representations (irreps). Each irrep
occurs with some degeneracy, giving a permutation factor,
which by Schur's lemma \cite{Simon96} is maximally-mixed.  Using the
expression for multiple copies of a qubit mixed state from
\cite{BBGMM05}, which gives the irreducible states of
$\rho_{p,q}^{\ox m}$ as a function of its Bloch vector and doing the
same for $Z^{\ox m} \rho_{p,q}^{\ox m}Z^{\ox m}$, we can compute
$S\left(\frac{1}{2}\rho_{p,q}^{\ox m}{+}\frac{1}{2}Z^{\ox m}
\rho_{p,q}^{\ox m}Z^{\ox m}\right)$ for $m$ up to several
hundred.

\begin{figure}[htbp]
\centering
\includegraphics[width=8.5cm]{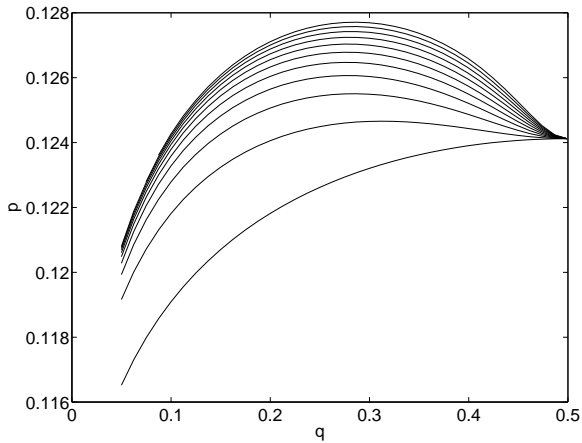}

\caption{Bit error rate $p$ at which the key rate goes to zero as a function of processing noise $q$ when
using various-sized repetition codes in the BB84 protocol. 
The curves are, from bottom to top, $m=1$,$m=10,20,\dots 100$,
illustrating the fact that a longer repetition code allows a higher threshold.  
As $m$ is increased, the optimal $q$ also grows.  Taking $m=400$ and $q=0.32$ gives
our best threshold of $0.1292$.}
\label{figure:NoiseVsThresh}
\end{figure}

In general, larger $m$ gives higher thresholds with the optimal $q\approx 0.3$ increasing slowly with $m$ 
(FIG \ref{figure:NoiseVsThresh}).  $m{=}400$ and $q{=}0.32$ give nonzero key rate up to $p{=}.1292$, but for larger $m$
the computation becomes quite slow.

{\em Discussion---}Given the pattern of improving thresholds with larger $m$, 
it is tempting to guess the best threshold within 
our family of codes will be when $m\rightarrow \infty$ as $q\rightarrow 0.5$.  
While we have not 
been able to do so, we hope that an asymptotic analysis of our 
key rates in the limit of large $m$ could be tractable.  Along these lines, 
note that an exact analysis of large repetition codes in the context of quantum
capacities was successfully carried out in 
\cite{SS96}.
     
We note that our codes are highly restricted, and it is not at all 
clear that they should be optimal.  One idea for better rates is to adapt the concatenation of repetition 
codes in conjugate bases used in \cite{DSS98,SmithSmo0506} to key generation, using a repetition code in the $X$ basis
to improve privacy amplification.  A more ambitious approach 
is to develop new degenerate codes for this problem, perhaps using the heuristic suggested in \cite{SmithSmo0506}.

The best upper bound on the BB84 key rate 
is $H(1/2{-}2p(1{-}p)){-}H(2p(1{-}p))$ \cite{AE}.  This gives an upper bound on the
threshold for BB84 of $p =  (1{-}1/\sqrt{2})/2 \approx 0.1464$, matching the bound due to the optimal
individual attack found in \cite{FGGNP97}.  There remains a significant gap between our lower bound of $0.129$ 
and this upper bound.

Our one-way protocols bear a striking resemblance to
two-way protocols using advantage distillation \cite{GL03}.  In
particular, an advantage distillation protocol can be described as
using a repetition code, with Bob sending the syndromes back to Alice.
Error correction and privacy amplification are performed on
blocks for which no error is detected, while the blocks for which an
error is detected are thrown away.  Without back
communication from Bob, Alice would not know the syndromes, and thus
be unable to discard blocks in which Bob had detected an error.
Our findings show that even in this case, with Alice ignorant of
the syndromes, and thus unable to discard bad blocks, there is still a
benefit in using a repetition code.  The repetition code
works ``better than expected'', because it collapses many phase errors
to a single logical phase error, while still providing information
about bit errors.  This benefit should also appear when the code is
used for advantage distillation with noisy processing.

One-way protocols with noisy
processing can be viewed quite naturally as distillation protocols for
twisted EPR pairs \cite{RS06,HHHO05}.  In \cite{RS06} it was
shown that noisy processing can be interpreted as the deflection of
Eve's correlations away from the sifted key into a ``shield'' system,
which purifies the noise added by Alice.  
Viewed in this way, the benefit of a repetition
code is that it allows us to combine the ``soft'' approach of
deflecting phase errors and the ``hard'' approach of correcting
\amplitude errors -- while learning about bit errors that we must
correct, we are simultaneously decreasing Eve's correlation with the
key, reducing the need for privacy amplification later.

{\em Acknowledgements---}We thank Debbie Leung, John Preskill, and Renato Renner 
for several valuable discussions.  This
work grew out of discussions between GS and JMR at the University of
Queensland, whose hospitality we appreciate.  JMR acknowledges
the Alexander von Humboldt Foundation, GS NSF grant
PHY-0456720 and Canada's NSERC, and JAS ARO contract
DAAD19-01-C-0056.

\end{document}